\documentclass[preprint,12pt]{aastex}
\usepackage{epsfig}

\newcommand\be{\begin{equation}}
\newcommand\ee{\end{equation}}

\newcommand{\mcommand}[1]{\ifmmode #1\else $#1$\fi}
\newcommand{\sci}[1]{\mcommand{\times 10^{#1}}}

\newcommand{\unit}[1]{\mcommand{\;{\rm #1}}}
\newcommand\meter{\unit{m}}
\newcommand\second{\unit{s}}
\newcommand\as{\unit{arcsecond}}
\newcommand\mas{\unit{mas}}

\newcommand\kpc{\unit{kpc}}
\newcommand\Mpc{\unit{Mpc}}

\newcommand\mps{\unit{m\;s^{-1}}}
\newcommand\cm{\unit{cm}}
\newcommand\km{\unit{km}}
\newcommand\kmps{\unit{km\;s^{-1}}}

\newcommand\gram{\unit{g}}
\newcommand\kg{\unit{kg}}

\newcommand\keV{\unit{keV}}

\newcommand\etal{et al.}

\newcommand\ltsim{\lesssim}
\newcommand\gtsim{\gtrsim}

\begin{document}

\title{Improving the Resolution of X-Ray Telescopes with Occulting Satellites}
\author{Craig J. Copi\altaffilmark{1}\email{cjc5@po.cwru.edu} \and Glenn D. Starkman\altaffilmark{1,2}\email{gds6@po.cwru.edu}}
\affil{
BOSS home page: \tt\url{http://boss.phys.cwru.edu/}}
\altaffiltext{1}{Department of Physics, Case Western Reserve
University, Cleveland, OH 44106-7079}
\altaffiltext{2}{Department of Astronomy, Case Western Reserve
University, Cleveland, OH 44106-7079}
\authoraddr{10900 Euclid Avenue\\ Cleveland, OH 44106-7079}

\shorttitle{Improving X-Ray Resolution with X-BOSS}
\shortauthors{Craig J. Copi and Glenn D. Starkman}

\begin{abstract}
One of the challenges of X-ray astronomy 
is how to both collect large numbers of photons yet 
attain high angular resolution. 
Because X-ray telescopes utilize grazing optics, 
to collect more photons requires a larger acceptance angle 
which in turn compromises the angular resolution.
All X-ray telescopes thus have angular resolution
far poorer than their diffraction limit.
Although collecting more photons is a desirable goal,
sometimes selective collecting fewer photons  may yield more information.
Natural (such as lunar) occultations 
have long been used to study sources on small angular scales.
But natural occulters are of limited utility because
of their large angular velocities relative to  the telescope,
and because of the serendipity of their transits.
We describe here how one can make use of an 
X-ray Big Occulting Steerable Satellite (X-BOSS) 
to achieve very-high resolution of X-ray sources.
An X-BOSS  could significantly improve the resolution of 
existing X-ray facilities such as the Chandra telescope, 
or X-ray Multiple Mirror (XMM) satellite, 
and could vastly improve the resolution of some future
X-ray telescopes, particularly Constellation X where sub-milliarcsecond
resolution is possible for a wide range of sources.
Similar occulting satellites could also be deployed in conjunction 
with planned space observatories for other wavebands.
\end{abstract} 


\slugcomment{\small Preprint: {\bf CWRU-P04-00}}

\section{Introduction}

One of the big challenges in doing X-ray astronomy
is the relatively low photon fluxes from  target sources.
The fact that X-ray mirrors operate only at grazing 
angles of incidence further exacerbates this problem.
Thus, while one might naively expect 
superb angular resolution from a $1.2\meter$ aperture X-ray telescope  
such as the one on board the Chandra satellite,
the $0.5$ arcsecond reality is far from the 
0.3 milliarcsecond nominal diffraction  limit,
and considerably worse than what is routinely achieved in
longer wavelength bands.  This situation is unlikely to 
change in the near future.  Indeed, current plans for
future X-ray missions opt for increased acceptance
angle (and thus increased photon count rate)  at
the price of reduced angular resolution.

But it {\em is} possible to achieve higher 
X-ray photon count rates and yet improve one's angular resolution.  
The necessary step is to separate the collection of photons 
from the means of achieving high resolution.
One way to do this is well-known---occultation.
When an astronomical body, such as the moon, transits the field
of view of a telescope, it occults different sources within the
field of view at different times.  By carefully measuring
the photon count rate as a function of time during the transit,
one can then reconstruct the projection of the surface brightness
in the field of view onto the path of the occulter.

Natural occulters {\em have} been used to achieve high-resolution
in X-ray observations; however, they have at least two distinct
disadvantages:
\begin{enumerate}
\item Although natural occultations can be predicted, they cannot be
  scheduled---target sources are therefore limited, and multiple
  occultations of the same source over the course of a few years
  are uncommon.
\item Natural occulters have large angular velocities relative to
  a telescope.  The shorter the transit time, the fewer photons one
  collects, and so the lower the resolution.  This is especially important
  for X-ray astronomy, where photon count rates are relatively low.
\end{enumerate}
There is however an alternative to natural occulters which can overcome
both of these disadvantages---a steerable occulting satellite.  Deployment
of large steerable occulting satellites has been discussed for optical and
near infra-red wavebands
\citep{Adams,Schneider,CS98,CS00},
mostly for
the purpose of finding planet around nearby stars, but also for
high-resolution astronomical observations.  However, such satellites are
naturally well-suited for observations in the X-ray and far-UV\@.  In the
longer wavelength bands, minimization of diffractive losses pushes one to
make the satellite as large as feasible, and deploy it as far as possible
from the telescope.  In the X-ray waveband one is far into the geometric
optic limit and diffraction of the X-ray photons around the satellite can
essentially be neglected; thus the optimal size and placement of the satellite 
are determined by one's ability to accurately position the satellite with respect to the
telescope-star line-of-sight and to minimize the satellite's velocity
perpendicular to that line-of-sight.  The resolution delivered by the
combination of the X-ray telescope and the X-BOSS is determined by the
collecting area of the telescope (and thus the photon count rate for a
source) and by the accuracy with which one can match the X-BOSS and telescope
velocity.  It is independent of the intrinsic resolution of the telescope.

For an X-ray telescope either in an eccentric high-Earth orbit (Chandra and XMM) or at the
L2 point of the Earth-Sun system (Constellation-X) we discuss in section
\ref{sec:location} where to position X-BOSS relative to the satellite.  In
section \ref{sec:blocking} of this letter we discuss the X-ray blocking
efficiency of a thick film and what it implies for the required thickness
of the occulter.  We also estimate in this section the required dimensions of an X-BOSS,
which are determined mostly by limitations on telemetry.
We discuss the steering of the X-BOSS in section \ref{sec:steer}.  
In section \ref{sec:resolve} we find the angular resolution that one
obtains as a function of X-BOSS-telescope relative angular velocity, 
and of photon count rate.  
In section~\ref{sec:skycoverage} we discuss the the sky
coverage that one could obtain in each location.  
Application of these techniques to specific sources is discussed briefly 
in section~\ref{sec:results}.  
Finally, section~\ref{sec:conclude} contains the conclusions.

\section{Locating an X-BOSS}
\label{sec:location}

The location of an X-BOSS is dictated by the location of the
telescope it is meant to occult.  The Chandra X-ray telescope is
in an elliptic orbit around the Earth with 
an apogee of  $145,417\km$ and a perigee of $16,026\km$.
The X-ray Multiple Mirror (XMM) Telescope will also be inserted into
an elliptic Earth orbit.
Other X-ray telescopes, such as Constellation X,  
may be located at the second Lagrangian point of the Earth-Sun system.  
The orbital issues are entirely
different for these two locations; 
we address each  in turn below.
Finally, some X-ray telescopes (Astro-E and XEUS) will be
placed in low earth orbit.  Because orbital velocities
are so high in low earth orbit, it is more difficult to make 
use of the approach we advocate here; we will not discuss
these further.

\subsection{Eccentric high Earth-orbit}
\label{subsec:orbitearth}

As described above, the Chandra satellite is in an eccentric high altitude
Earth orbit.  The period of this orbit is 64 hours.  The satellite
therefore has an average angular velocity of about 6 arcseconds per second.  
An occulting satellite leading or following in Chandra's orbit would transit a source at
approximately that rate.  The planned X-ray Multiple Mirror Telescope (XMM)
has a similar orbit with a shorter 48 hour period.
Given that the attainable angular resolution is related to the angular 
velocity of transit, 
the resolution that one could achieve with these orbits is minimal.

A great improvement is to place the X-BOSS in an orbit identical to that of
the telescope but slightly modified by shifting the apogee and perigee, by
changing the phase of the satellite in the orbit, or by rotating the orbit.
In all cases these modifications will put X-BOSS in an orbit with the same
period as telescope.  In such orbits the velocity perpendicular to the line
of sight of X-BOSS and the telescope can be quite low.  For example, consider
placing X-BOSS in an orbit identical to that of the telescope (in terms of
apogee, perigee, and orbital phase) but rotated about an axis through the center of the Earth
in the plane of the orbit and perpendicular to the line connecting apogee and perigee.  
The component of the relative velocity between X-BOSS and the telescope
perpendicular to the line-of-sight between them is then zero throughout the
entire orbit.  Unfortunately, such an orbit intersects the telescope orbit at
two points with disastrous consequences.  The other orbital modifications
mentioned above can alleviate this problem by enforcing a minimum
separation of, for example, $10\km$ between the two spacecrafts.  Although
$10\km$ may seem fairly close, note that each spacecraft is only a
few to tens of meters across; random errors therefore have a probability
less than $10^{-9}$ per orbit crossing of causing catastrophic failure.
The importance of utilizing these orbit modifications is explained more
fully in sections~\ref{sec:steer} and \ref{sec:skycoverage}.

\subsection{Orbit at L2}
\label{subsec:orbitL2}

We have previously discussed the orbital advantages of placing a large
occulter at L2 in greater detail \citep{CS00}.  Here we will
highlight the important points.
Orbits around L2, both in the plane of the ecliptic and oscillations
perpendicular to this plane, have periods of about 6 months independent of
their distance from L2 (for distances $\ltsim 10^4\km$).  Therefore the
local gravity is very small.  Both the total velocity and acceleration of
orbits around L2 (relative to the L2 point) are on par with those we might attain
through carefully 
tuning the orbit of X-BOSS relative to that of Chandra or XMM; 
the relative velocity of the satellite and telescope due to the
motion of L2 about the Sun is of the same magnitude.  If corrections are
made to the X-BOSS orbit to cancel these then the
acceleration perpendicular to the line-of-sight is about
$5\sci{-10}\meter\second^{-2}$.  Thus if the perpendicular velocity of
X-BOSS relative to a particular line-of-sight between the telescope and some source
is canceled by firing rockets, the perpendicular velocity will remain less than $10^{-4}\mps$ for 
at least a day.  Tuning the velocity of X-BOSS, therefore, can be done very 
easily at L2.

\section{Making an X-ray Occulter}
\label{sec:blocking}

\subsection{Thickness}

\begin{figure}
  \leavevmode\center{\epsfig{figure=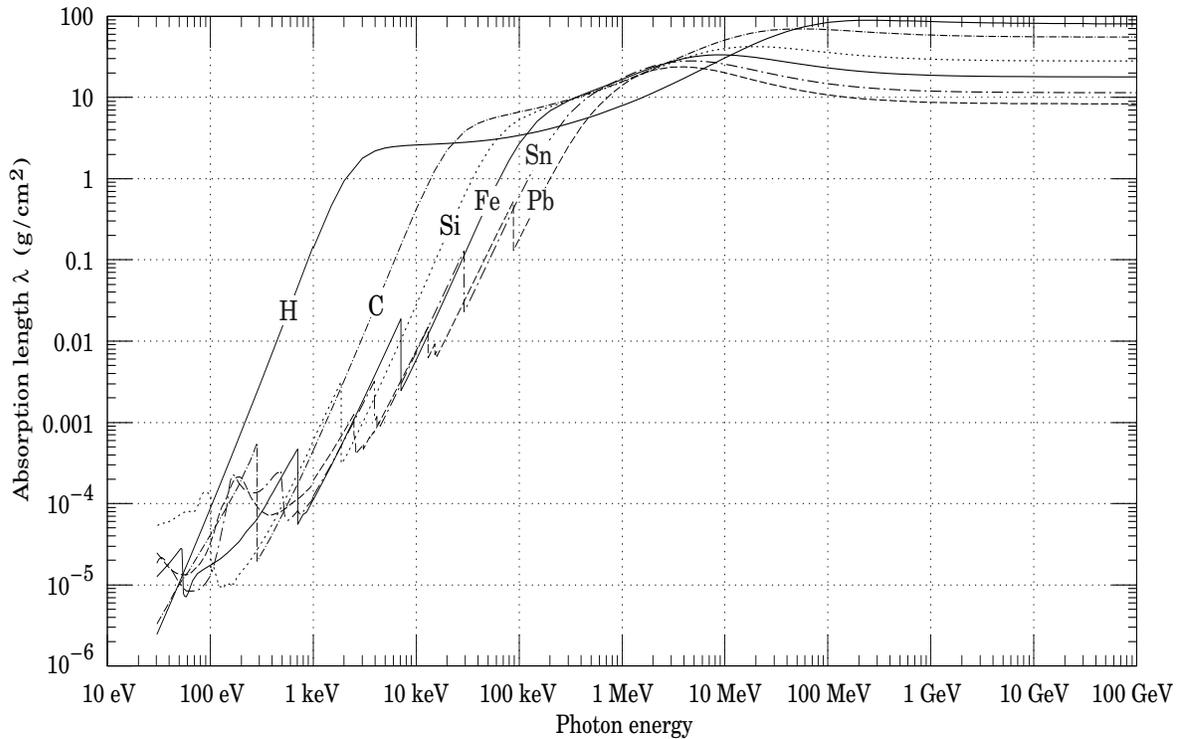,height=5 in,width=8.5 in}}
  \caption{The photon mass attenuation length $\lambda=1/(\mu/\rho)$
for  various elemental absorbers as a function of photon energy
($\rho$ is the density).  The figure is obtained from the particle
data book,  figure 23.11 (http://pdg.lbl.gov).  
The data for $30{\rm eV} < E < 1 {\rm keV}$ are obtained from 
http://www-cxro.lbl.gov/optical\_constants 
(courtesy of Eric M. Gullikson, LBNL).
The data for $1{\rm keV} < E < 100 {\rm GeV}$ are from 
\protect\url{http://physics.nist.gov/PhysRefData}, thru the courtesy of 
John H. Hubbel (NIST).
}
\label{fig:photonattenuation}
\end{figure}

The attenuation length of X-ray photons  in elemental matter
is shown in figure~\ref{fig:photonattenuation}.
Except in hydrogen, it is approximately $3\times 10^{-4} \gram \cm^{-2}$ at
$1\keV$, and $10^{-2} \gram\cm^{-2}$ at $10\keV$.  
Thus a square $10\meter$ on a side and one attenuation length
thick has a mass of $0.3\kg$ at $1\keV$, and $10\kg$ at $10\keV$.
At a typical density of $3 \gram \cm^{-3}$, these represent thicknesses
of just $1$ micron  and  $30$ microns respectively.

A useful occulter would need to be $3\hbox{--}5$ attenuation lengths thick,
and so $3\hbox{--}5$ microns and $1\hbox{--}2\kg$ to operate at $1\keV$,
and $100\hbox{--}150$ microns and $30\hbox{--}50\kg$ to operate at $10\keV$.
Even at $100\keV$ a $10\meter\times10\meter$ lead film 
$0.6\unit{mm}$ thick at a mass of $600\kg$ would provide 3 attenuation 
lengths of occultation.

If positioning technology improved to the point where one could
reduce the size of the occulter to $1\hbox{--}2\meter$, then even gamma ray
occulters would be of reasonable mass.

\subsection{Size}

The size of the occulting satellite depend  on
two factors---the aperture of the telescope and 
the accuracy with which one can position the occulter.  

The apertures of typical X-ray satellites are about $1\meter$.
This sets a lower bound on the dimensions of the occulter.
Once the occulter is larger than the aperture of the X-ray telescope,
there is essentially no effect on resolving power.

Next we will estimate how well we can determine the position of X-BOSS 
in the plane perpendicular to the telescope-source line-of-sight 
it is meant to occult.  
Consider a telescope separated from the X-BOSS by a distance $r$.  
We can mount a small diffraction-limited optical  telescope of diameter $d$ 
on the underside of the occulter.  Using this telescope
we can establish the relative positions
of the two satellite  to within approximately
\be 
\delta x = 1.2 r \frac{\lambda}{d}  
= 0.5 \meter \frac{r}{1000\km} \frac{\lambda/400\unit{nm}}{d/1\meter}
\ee
A $1\meter$  positioning accuracy therefore requires
a $50\unit{cm}$ finder scope at $1000\unit{km}$ separation,
proportionately smaller at smaller separations.  
This is quite feasible, especially since the finder scope 
need not have a full UV plane.

An important question is  whether one will collect
enough photons to reach the diffraction limit of the
angular resolution.   There are two principal
options---rely on reflected sunlight or  
shine a laser from the X-ray telescope onto the X-BOSS telescope.
Columnation is not a significant problem, 
as seen by our calculation of the diffraction limit above.  
However,  sunlight has an intensity of $1000\unit{W}\meter^{-2}$, 
which will be difficult to match with a laser anyway. 
Assuming isotropic scattering from the telescope,
and a total reflecting area of $1\meter^{2}$,
this results in a flux at the X-BOSS  of $3\sci{8}\second^{-1}$,
at $1000\km$ (falling as $1/r^2$).  Detailed studies
of existing telescopes (Chandra, XMM) would be required
to precisely quantify our ability to  locate the
telescope relative to the X-BOSS, however, these
estimates suggest that determining the relative
position to within $1\hbox{--}3 \meter$ is not unrealistic.
In the case of yet-to-be launched telescopes, 
the mounting of a small reflector on
one or more corner of the telescope would be of
definite benefit.

Although we have argued that we can determine the
relative position of an X-BOSS and an X-ray telescope 
to within about a  meter, we must also be able to reduce
the velocity to a fraction of a meter per second.
This can be done by a simple bootstrapping procedure.
Two position determinations each with error of $\Delta x$,
made a time $t$ apart, determine the velocity within
$\Delta v \simeq \sqrt{2} \Delta x/t$ (assuming the
error in $t$ to be negligible).  If the relative velocity
can  be canceled within errors by accurately firing rockets,
then  the ability to reduce $\Delta v$ is limited by 
the time one can allow between position determinations, $t=\Delta x/\Delta v$.
This time is limited by the orbital accelerations, 
but is thousands of seconds for the elliptic earth orbits  of interest
(cf.~subsection \ref{subsec:earthorbit})
and hundreds of thousands of seconds for orbits at L2.
(cf.~subsection \ref{subsec:L2orbits}).
In practice it may be desirable to gradually
reduce the relative velocity using repeated position
determinations and rocket firings.

\section{Steerability}
\label{sec:steer}

In order to successfully resolve objects it will be necessary
to frequently change the velocity of the satellite. 
These velocity changes will occur for  two principal reasons:
to move from one target source to another,
and to match the velocity of the X-BOSS to that of the X-ray telescope.
While solar radiation pressure might be used to some advantage,
it will be necessary to make some velocity adjustments using rockets.
The number and size of such adjustments may be the limiting factor
on the useful lifetime of the X-BOSS.

A change $\Delta v$ in the satellite's velocity  is related
by momentum conservation to the mass of propellant ejected,
$\Delta m_{\rm propellant}$, and the velocity of ejection $v_{\rm ejection}$:
\be
\Delta v_{\rm sat}
= \frac{\Delta m_{\rm propellant}{ v}_{\rm ejection}}{m_{\rm sat}} .
\ee
If $N$ is the number of desired major rocket-driven velocity changes,
then we must keep
$(\Delta m_{\rm propellant}/m_{\rm sat})\leq N^{-1}$.
(The mass of propellant ejected will of course vary on the particular
maneuver, but here $\Delta m_{\rm propellant}$ is taken to be some 
typical mass of propellant expended per orbit reconfiguration.)
We therefore can accommodate only a limited number of such rocket firings:
\be
N \leq \frac{v_{\rm ejection}}{\Delta{ v_{\rm sat}}} .
\ee
Off-the shelf, low-cost ion engines are currently available
with ejection velocities of $20 \kmps$, and
more expensive systems with $30 \kmps$ performance 
have been developed, thus
\be
N \leq \frac{30\kmps}{\Delta{ v_{\rm sat}}} .
\label{eqn:Ncorrect}
\ee

Consider first the need to match the velocities of the two  spacecraft
so that a long occultation can occur.
$\Delta{ v_{\rm sat}}$ is then the relative velocity 
of the X-BOSS and the telescope in their orbits.
In determining the sky coverage for elliptic Earth orbits 
in section~\ref{subsec:earthorbit} above, 
we have considered only orbital configurations with relative velocities 
between the telescope and the  X-BOSS of less than $10\mps$.
(Near L2, the relative velocities of relevance are typically 
considerably smaller than that.)
If $\Delta{ v_{\rm sat}}\simeq 10\mps$, then this implies 
$N\leq 3000$,
which is a reasonable  quota of corrections for a mission with a 3--5 year
lifetime, 
given the typical 2--3 day orbital period of Earth-orbiting
X-ray telescopes. 

The second type of velocity correction that will be required
is target acquisition---the readjustment of the orbit
of the occulter so as to allow the occultation of a new 
target source. 
For satellites separated by $1000\km$ near L2, relative velocities
are only $v_{\rm sat} = {\cal O}(10^{-4}\kmps)$,
and the expression for
$N$ (equation~\ref{eqn:Ncorrect}) shows that any constraint on target
choice or order does not come from concerns about conserving
propellant.
For telescopes in orbit about the  Earth, 
the matter is quite different.
Here orbital velocities are $v_{\rm sat} = {\cal O}(1\kmps)$,
and so it is clear from the allowed number of orbital
corrections~(\ref{eqn:Ncorrect}) that
one cannot indiscriminately rocket from one target to another on the sky.
One solution might have been to sail in the solar radiation pressure.
However, the solar radiation pressure is approximately $P_{\rm
  solar}=6\times10^{-6}\unit{Pa}$. 
For an areal density of just $1.5\times 10^{-3}\gram\cm^{-2}$
(five attenuation lengths as $1\keV$),
this results in an acceleration of only $4\times10^{-4}\meter\second^{-2}$.
At this rate it takes  about a month to change velocity by $1\kmps$.

Clearly one cannot reposition randomly on the sky.  
However the velocity difference between two orbits which
result in occultation of target sources one degree apart are only
of order $15\mps$.  Solar sailing can cause velocity changes
of this order in under a  day.  
Moreover, the allowed number of orbital corrections~(\ref{eqn:Ncorrect})
indicates that rocket driven
corrections of this magnitude can be made of order 1000 times.
How many corrections we can make, and how many sources we can
therefore target for occultation,
clearly depends on exactly how we use the satellite.  
A reasonable program of observations certainly seems possible.

\section{Resolution}
\label{sec:resolve}

When used in conjunction with an X-BOSS the telescope acts as a
light bucket.  The angular resolution of the telescope itself is
irrelevant; instead the collecting area is the important telescope
parameter.  The angular resolution of the system will come from probing the 
lightcurve as X-BOSS transits a source.

\begin{figure}
  \leavevmode\center{\epsfig{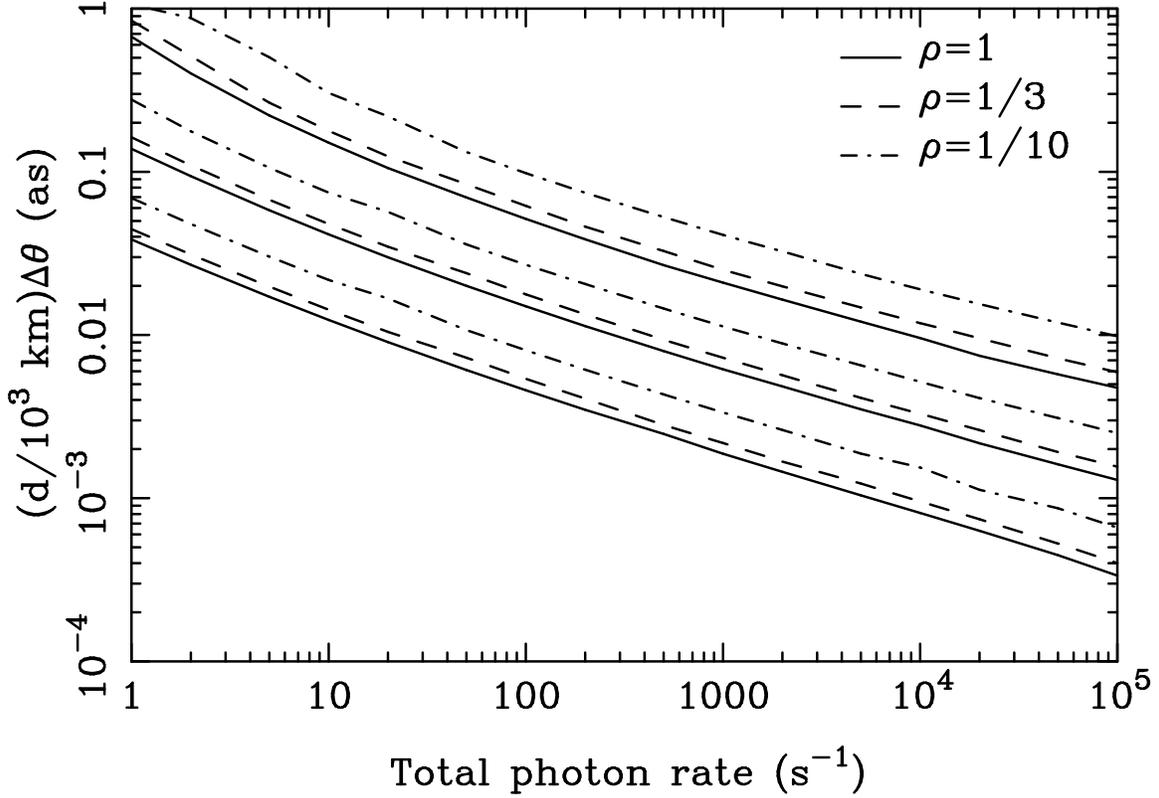}}
  \caption{The minimum angular separation of two X-ray sources resolvable
    at the 95\% confidence level.  The limits are shown for intensity
    ratios, $\rho=1$ (solid),
    $1/3$ (dashed), and $1/10$ (dashed-dotted).  The upper set of three curves
    are for $\mu_\perp = 10\mas\second^{-1}$, the middle set of three curves are
    for $\mu_\perp = 1\mas\second^{-1}$, and the lower set of three curves
    are for $\mu_\perp = 0.1\mas\second^{-1}$.  Note that the total photon
    rate is of photons detected in the telescope (without the presence
    X-BOSS), not photons incident on the telescope. Here $d$ is the
    distance between the telescope and X-BOSS in units of $10^3\km$.
    }
  \label{fig:resolution}
\end{figure}

To study the angular resolution of X-BOSS we consider the simple case of
identifying a binary source. Let $f (\vec x, t)$ be the normalized
lightcurve (number of photons detected per second) generated as X-BOSS
scans across a single source at a position
$\vec x$ in the plane of X-BOSS\@. The lightcurve is the number of photons
detected as a function of time.  It is normalized such that the value is
one (in the detector) when X-BOSS is not present. Since X-rays have
extremely short
wavelengths we can approximate the diffraction pattern produced by the
satellite simply by the geometric shadow projected on the telescope.  This
reduces the lightcurve to a calculation of the area of the telescope not
under the shadow of the occulter.  We write the lightcurve of a single
source as
\be 
l_1 (\vec x, t) = I_1 f (\vec x, t)
\ee
and the total lightcurve for two sources at $\vec x_1$ and $\vec x_2$
can be written as
\be
l_2 (\vec x_1, \vec x_2, t) = I_2 \left[ \rho f (\vec x_1, t) +
  (1-\rho) f (\vec x_2, t)\right].
\ee
Here $I_i$ is the total intensity of the system for $i=1$ or $2$ sources
and $\rho$ is
the intensity 
ratio of the two sources.  We would like to find the minimum separation of
two sources that can be distinguished from a single source.  An observation
consists of a sequence $\left\{O_j (\vec x)\slash j=1,...,n \right\}$
of measurements of the integrated lightcurve between 
times $t_{j-1}$ and $t_j$:
\be 
O_j (\vec x) = \int_{t_{j-1}}^{t_j} dt\; l_i (\vec x, t).
\ee
To obtain
limits on the minimum separation we first evaluate the number of photons
expected between time $t_0$ and $t_k$
\be
L_{i,k} (\vec x) \equiv \int_{t_0}^{t_k} dt\;l_i (\vec x, t) =
\sum_{j=1}^k O_j (\vec x)
\ee 
where $i$ is 1 or 2 as above.
Assuming the counts in each time bin, $[t_{j-1},t_j)$, are Poisson distributed the
likelihood of a model with $i$ sources given an underlying model with 2
sources is 
\be 
{\cal L}_i = \prod_{k=1}^N \left. e^{-L_{i,k}} \left( L_{i,k} \right)^{L_{2,k}}
\right/ \left( L_{2,k}\right)!.
\ee
Finally the quantity
\be 
T = -2\log \left (\frac{{\cal L}_1}{{\cal L}_2} \right)
\ee
is $\chi^2$ distributed with 4 degrees of freedom ($t_0$, $x_2-x_1$, $I$,
and $\rho$) and allows us to
calculate the probability of misidentifying a binary source as a single source.
This probability depends on 
$\mu_\perp$, the angular velocity of X-BOSS as it transits the source.
The results for the 95\% confidence limits as a function of the intensity
in a $1.2\meter$ aperture telescope for $\rho=1$, $1/3$, and $1/10$ and for
$\mu_\perp=10\mas\second^{-1}$, $\mu_\perp=1\mas\second^{-1}$, and
$0.1\mas\second^{-1}$ are shown in figure~\ref{fig:resolution}.  In
producing figure~\ref{fig:resolution} we assumed a uniform response over
the surface of the telescope.  A more complicated response function may
improve resolution slightly.

The simple analysis employed here uses the edges of X-BOSS in a single
occultation.  In practice it would be necessary to obtain multiple
projections to resolve a source in two dimensions.  This could be facilitated
by putting slits at various angles in X-BOSS that allow for sources to be
occulted by different regions of the satellite in different ways during a
single transit.

\section{Sky Coverage}
\label{sec:skycoverage}

The issues of resolution and sky coverage are closely related.  Here sky
coverage is the fraction of the sky for which a particular angular
resolution can be obtained.  While one can reposition X-BOSS to be in an
arbitrary direction on the sky relative to the X-ray telescope, this
frequently leads to large relative velocities and accelerations between the
occulter and telescope perpendicular to the line-of-sight, thus leading to
poor resolution (see figure~\ref{fig:resolution}).  Conversely, extremely
good resolution is possible if the relative velocity during the occultation
is kept quite low; however this requires either special orbits (and thus
very little sky coverage) or expenditures of fuel.  Here we will explore
the sky coverage that can be obtained subject to a number of constraints.

\subsection{Elliptic Earth Orbits}
\label{subsec:earthorbit}

We consider placing an X-BOSS in orbit around the Earth with nearly the
same orbital parameters as an X-ray telescope.  As discussed in
section~\ref{subsec:orbitearth} we then allow for small alterations in the
X-BOSS orbit which change the direction of the line-of-sight from the
telescope to X-BOSS while keeping the X-BOSS period fixed.
Our first constraint is that the
minimum separation of the X-BOSS and telescope in their orbits must be larger than $10\km$.  
(If this safety factor can be reduced then greater sky coverage may be possible.) 
We then follow the two spacecrafts in their orbits to see what sky coverage 
these orbits afford.
To limit the expenditure of propellant we 
consider making observations only when the relative velocity of the
two satellites perpendicular to the line-of-sight is sufficiently small,
here we require $v_\perp^{\rm orb} <
10\mps$ (see section~\ref{sec:steer}).  Prior to an observation this
velocity can be reduced to the desired range by firing the X-BOSS rockets
(a small correction requiring acceptable use of consumables).

After maneuvering to correct the perpendicular relative velocity between the
telescope and X-BOSS, they would still be accelerating relative to each
other during the observation.  The provides one limit on the total transit
time of the observation.   There would also be some residual error in the
velocity correction, leaving X-BOSS with a component of its velocity perpendicular to the
line-of-sight.  This provides another limit on the total transit time of the
observation. Combining these two, the total transit time over which the
observation can made is
\be
t_{\rm obs} = \min \left(\sqrt{\frac{2w}{a_\perp^{\rm orb}}},
  \frac{w}{\mu_\perp^{\rm max} d} \right).
\ee
Here $a_\perp^{\rm orb}$ is the perpendicular linear acceleration, $w$ is the
width of X-BOSS, $d$ is the distance between the two spacecrafts, and
$\mu_\perp^{\rm max}$ is the maximum angular velocity allowed to obtain a
particular resolution.
If we require that the angular velocity perpendicular to the line
of sight at the end of the observation be less than the same maximum value,
$\mu_\perp^{\rm max}$, then we obtain the constraint
\be
a_\perp^{\rm orb} t_{\rm obs} < \mu_\perp^{\rm max} d.
\label{eqn:constraint1}
\ee
Of course when we cancel $v_\perp^{\rm orb}$ before the observation we are
also making a small change to the orbit.  This leads to an extra
acceleration that must also be small.  If we require that this acceleration
also not produce a large final velocity we obtain the constraint
\be
\frac{v_\perp^{\rm orb} t_{\rm obs}^2}{r^3 d} < \frac{\mu_\perp^{\rm
    max}}{2g R_\oplus^2},
\label{eqn:constraint2}
\ee
where $r$ is the distance from X-BOSS to the (center of the) Earth,
$g=9.8\meter\second^{-2}$, and $R_\oplus$ is the radius of the Earth.

The sky coverage on each change of X-BOSS orbit is not large. To increase
the amount of sky
accessible to observation we consider moving X-BOSS between orbits that are
similar to the orbit of the X-ray telescope.  Throughout we will consider
modifications of the X-BOSS orbit that leave the period unchanged.  Over
many orbits this is a 
desired feature since it prevents the times at which X-BOSS and the telescope
achieve apogee and perigee from drifting apart, requiring a
large expenditure of fuel to correct.  The orbital modifications we
consider are increasing or decreasing the apogee distance (while preserving 
the semi-major axis and thus the period), rotating the orbit
about all three axes, and introducing a phase shift (time of apogee) into the
orbit.  For this study we taken the X-ray telescope to be the Chandra
satellite and allowed changes in apogee (and perigee) of
$\pm200\km$, rotations about the two axes in the plane of the orbit of
$\pm1^\circ$, rotations in the plane of the orbit of $\pm0.4^\circ$, and
time shifts of $\pm100\second$.  All of these changes are relative to
Chandra's orbit.  These changes can be accomplished using ion engines
several thousand times before exhausting the supply of expendables (see
section~\ref{sec:steer}).  Since occultations are best done near apogee and 
1000 or so orbital periods is the Chandra mission lifetime, this rate of
consumption of expendable is acceptable.

Using Monte Carlo techniques,
we studied the orbits in this region of parameter space 
subject to two constraints: the minimum separation of the X-BOSS and telescope 
must be at least $10\km$, and somewhere in the orbit the
perpendicular velocity must be less than $10\mps$.  We generated $100,000$
orbits that satisfy these criteria.  Next, for a variety of photon count rates
and desired resolutions we used the resolution results show in 
figure~(\ref{fig:resolution}) to determine $\mu_\perp^{\rm max}$.  Finally
we checked which lines-of-sight satisfied the velocity and acceleration
constraints~(\ref{eqn:constraint1}, \ref{eqn:constraint2}) and thus which
parts of the sky can be observed.  The results are shown in
figure~\ref{fig:skycoverage} assuming the width of X-BOSS is $10\meter$. 
Here even for very intense sources (I=$10^5\second^{-1}$) only 20\%
of the sky can be covered with a a resolution of
$\Delta\theta=0.1\as$.  This tight constraint is due principally to the fact
that X-BOSS is accelerating during the time that both of the sources are
occulted leading to a large velocity by the end of the observation which
occurs when the transit is complete.  
To counteract this we could use a narrower satellite, 
use a satellite with slits in it so that we do not have to wait until the far
edge starts unocculting the sources, or fire the rockets while both sources
are occulted to cancel the velocity.  To model these possibilities we have considered 
a satellite that accelerates over only $2\meter$ between the onset and end of
a transit.  The results are shown in figure~\ref{fig:skycoverage}b.  
Here we see a tremendous improvement in sky coverage and resolution.  
For intense sources, $I=10^5\second^{-1}$, we can obtain
$\Delta\theta=0.1\as$ over 
40\% of the sky and $\Delta\theta=0.02\as$ over 20\% of the sky.
Larger sky coverages would be obtained if we relaxed the criteria on the 
orbital velocity difference between the Chandra and X-BOSS orbits.
This would be justified if the propellant velocities of ion engines
rose about $30\kmps$, or if we could make do with a smaller number of orbital 
corrections.

\subsection{L2 Orbits}
\label{subsec:L2orbits}

As discussed above (section~\ref{subsec:orbitL2}) the orbits at L2 are much 
simpler to manage than orbits around the Earth.  Full sky coverage can be
obtained and the velocity perpendicular to the line-of-sight can be chosen
as desired.  Figure~\ref{fig:resolution} best represents what can be
achieved at L2.  Since the perpendicular velocity can be chosen, extremely
high angular resolution is possible.  Even sub-milliarcsecond resolution is
possible for many sources ($I\gtsim 4\sci{3}\second^{-1}$) when
$\mu_\perp=0.1\mas\second^{-1}$.

\section{Results}
\label{sec:results}

This is an exciting time for X-ray astronomy.  Two new X-ray telescopes
(the Chandra Advanced X-ray Astronomical Facility, and the X-ray Multiple
Mirror telescope (XMM)) have been successfully launched, while another
(Astro-E) is being readied for launch.  Of these three, Chandra and XMM are
in highly elliptical high altitude earth orbits
(cf.~Table~\ref{tab:xray-telescopes}) while Astro-E is headed for a
circular low earth orbit.  In addition, at least two major X-ray
space-observatories are being planned: Constellation X, with launch
scheduled for 2003, and XEUS with a target date of 2007.  Constellation X
will be placed at the L2 point of the Earth-Sun system, while XEUS, like
Astro-E, will be placed in low Earth orbit.

While this may seem a remarkable proliferation of X-ray telescopes, each
mission has its own emphasis.  In building an X-ray telescope there is a
direct competition between large effective area (and thus sensitivity) and
small acceptance angle (and thus high angular resolution).  Therefore one
must choose whether to build an instrument which aims for high
angular-resolution or one which has a goal of achieving high sensitivity.
Chandra is the only high angular resolution instrument of the listed
missions, with a maximum resolution of $0.5\as$ and thus has the relatively
small effective area given above~(\ref{eqn:AChandra}).  The other
instruments all aim for large effective area, and so sacrifice angular
resolution.  XMM, which is already flying, has considerably larger
effective area than Chandra (and thus much lower angular resolution).
Astro-E will have even larger effective area.  Constellation-X will consist
of multiple X-ray telescopes flown in formation, with a total effective
area considerably greater than either XMM or Astro-E; it too has relatively
low angular resolution compared to Chandra.  Finally XEUS will have a huge
effective area, and will be designed to be expandable.  Its angular
resolution is better than XMM, Astro-E, or Constellation-X but still not as
good as Chandra.  The properties of the existing and planned
telescopes are shown in Table~\ref{tab:xray-telescopes}.

Throughout we have considered the photon rate in the detector, not at the
surface of the telescope.  An X-ray telescope has an effective area, $\cal
A$, which includes the geometric collecting area (since grazing optics are
used the collecting area is not the full beam) and the efficiency of the
X-ray detector.  As an example, with Chandra
\be 
 {\cal A}_{\rm Chandra} \approx 700 \cm^2
 \label{eqn:AChandra}
\ee
for $E\approx 1\keV$.  This is the area to be used as the area of the
telescope, not the geometric area as in the case of optical telescopes.
The effective area for existing and planned X-ray telescopes is given in Table~\ref{tab:xray-telescopes}.

The luminosity of X-ray sources varies greatly.  Black holes in the cores
of nearby galaxies have 
\be 
{\cal L}_{\rm bh} \approx 10^{38\hbox{--}40}\unit{erg}\second^{-1} =
6.2\sci{46\hbox{--}48} \keV\second^{-1}
\ee
in the $0.2\hbox{--}2.4\keV$ energy range.  This leads to a photon rate at
the surface of the detector of
\be
\Gamma_{\rm bh} = (0.052\hbox{--}5.2)\sci{-2} \left (\frac{E}{\keV}\right) 
\left ( \frac{d}{1\Mpc}\right)^{-2}
\left (\frac{\cal A}{\cm^2}\right) \second^ {-1},
\ee
where the energy, $E$, we observe at is given in keV and $\cal A$ is the
effective area of the X-ray telescope as discussed above.

An active galactic nucleus (AGN), Seyfert galaxy, or the core of X-ray
clusters can be much more luminous
\be 
{\cal L}_{\rm AGN} \approx 10^{40\hbox{--}44}\unit{erg}\second^{-1} =
6.2\sci{48\hbox{--}52} \keV\second^{-1}.
\ee
However, since they are approximately $100\Mpc$ away the photon rate 
is only 
\be
\Gamma_{\rm AGN} = 5.2\times (10^{-6}\hbox{--}10^{-2})\left
  (\frac{E}{\keV}\right) 
\left ( \frac{d}{100\Mpc}\right)^{-2}
\left (\frac{\cal A}{\cm^2}\right) \second^ {-1}.
\ee
Galactic microquasars are somewhat less luminous
\be 
{\cal L}_{\rm microquasar} \approx 10^{39}\unit{erg}\second^{-1} =
6.2\sci{47} \keV\second^{-1},
\ee
since they are in our own galaxy, though, the photon rate is fairly high
\be
\Gamma_{\rm microquasar} = 52 \left (\frac{E}{\keV} \right) 
\left ( \frac{d}{10\kpc}\right)^{-2}
\left (\frac{\cal A}{\cm^2}\right) \second^ {-1}.
\ee

\begin{figure}
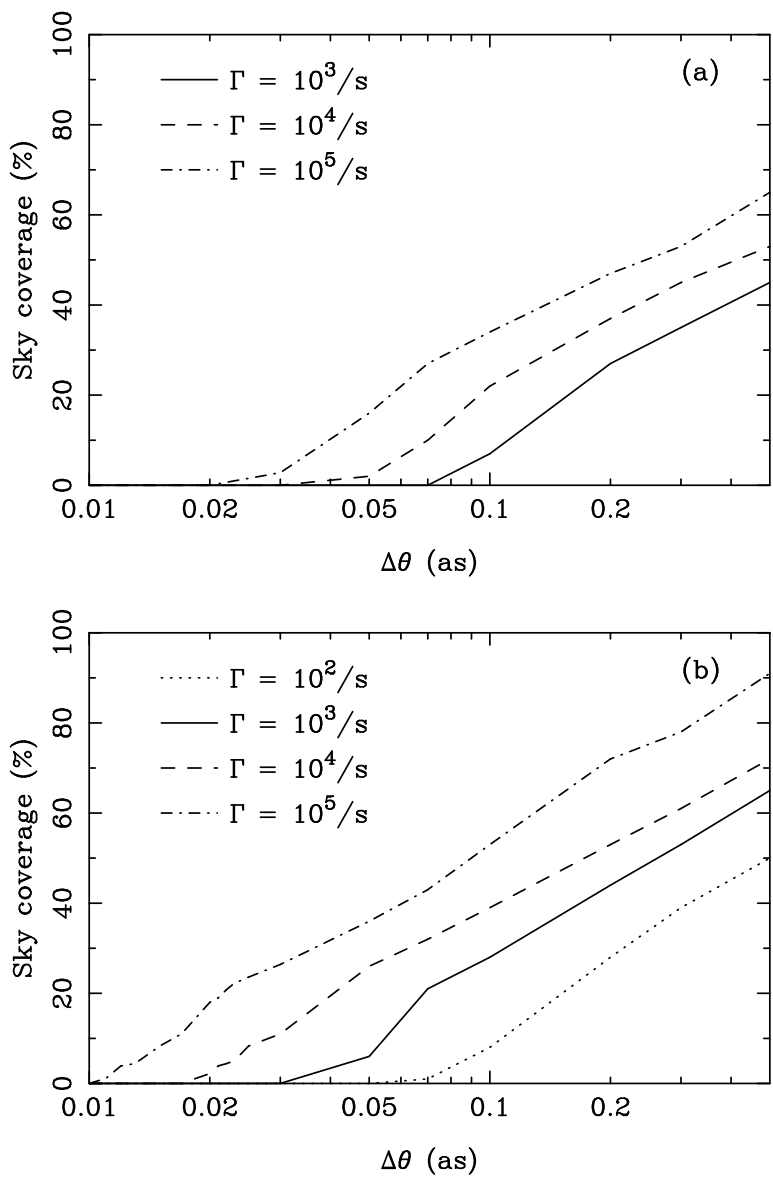

  \leavevmode\center{\epsfig{figure=skycoverage_10m.ps,height=4in,angle=-90}}
  \leavevmode\center{\epsfig{figure=skycoverage_2m.ps,height=4in,angle=-90}}
  \caption{The fraction of the sky that can be observed as a function of
    the desired resolution, $\Delta\theta$, and the photon rate in the
    detector (as in figure~\protect\ref{fig:resolution}), $\Gamma$, for
    an X-BOSS used in conjunction with Chandra.  (a) A satellite
    width of $10\meter$ is assumed here. (b) A satellite
    width of $2\meter$ is assumed here.  A larger satellite can obtain
    these results if velocity corrections are made during the observation.
    See the text for details.
    }
  \label{fig:skycoverage}
\end{figure}

For a $10\meter$ X-BOSS employed in conjunction with Chandra we find
(figure~\ref{fig:skycoverage}a) that for the brightest sources (galactic 
microquasars) we can obtain $\Delta\theta=0.5\as$ over about 30\% of the
sky with the sky coverage falling quickly until at $\Delta\theta=0.1\as$
very little of the sky is accessible.  This represents a modest gain over
what can be obtained by Chandra without the aid of X-BOSS.  For a $2\meter$ 
X-BOSS (figure~\ref{fig:skycoverage}b) the situation is much
better.  Here $\Delta\theta=0.5\as$ can be obtained over about 50\% of the
sky, $\Delta\theta=0.1\as$ over about 20\% of the sky, and
$\Delta\theta=0.05\as$ over about 5\% of the sky.  Thus significant
improvements are attainable through the use of an X-BOSS with Chandra.

\begin{figure}
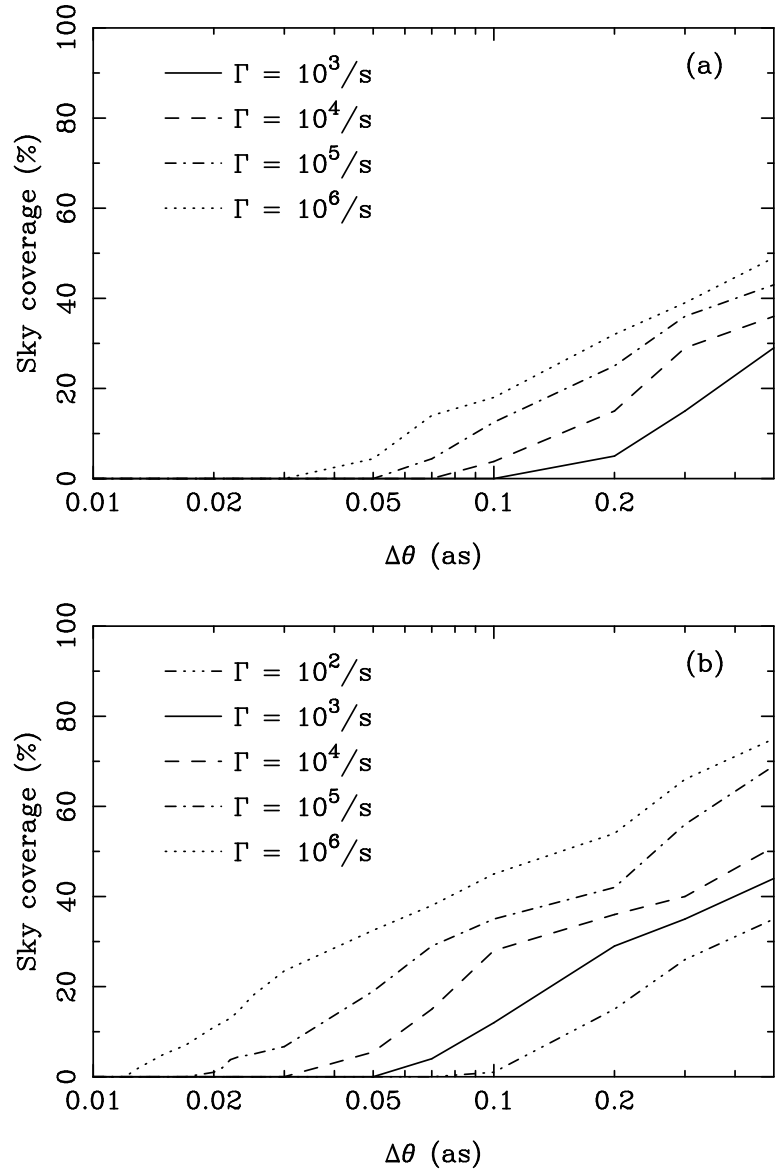

  \leavevmode\center{\epsfig{figure=skycoverage_XMM_10m.ps,height=4in,angle=-90}}
  \leavevmode\center{\epsfig{figure=skycoverage_XMM_2m.ps,height=4in,angle=-90}}
  \caption{
    The same as figure~\protect\ref{fig:skycoverage} for an X-BOSS used in
    conjunction with XMM\@.
    }
  \label{fig:skycoverageXMM}
\end{figure}

For an X-BOSS employed in conjunction with XMM the situation is similar.
Although XMM has a shorter period than Chandra its has an effective area about
$3$ times larger (Table~\ref{tab:xray-telescopes}).  For a $10\meter$
X-BOSS (figure~\ref{fig:skycoverageXMM}a) the skycoverage that can be
obtained for each incident photon rate, $\Gamma$, is less than can be
obtained by Chandra (compare to figure~\ref{fig:skycoverage}a) even with
the factor of $3$ increase in effective area that XMM provides.  For a
$2\meter$ X-BOSS (figure~\ref{fig:skycoverageXMM}b) the sky coverage for
XMM and Chandra are closer though Chandra is still superior.  Note that for
both sizes of X-BOSS tremendous improvements over the nominal $15\as$
resolution for XMM are obtained.

At L2 the situation is even better.  Since we can tune the velocity
relative to the line-of-sight more easily, great improvements in resolution 
are readily available (figure~\ref{fig:resolution}).  Sub-milliarcsecond
resolution can be obtained for sources with photon rates $\Gamma\gtsim
1000\second^{-1}$. For a single Constellation X modules, which will have an 
effective area of about $15,000\cm^2$, the brightest AGN's, X-ray cluster
cores, and galactic black holes will have $\Gamma\approx 800\second^{-1}$
we can obtain $\Delta\theta\approx 2\mas$.

\section{Conclusions}
\label{sec:conclude}

We have found that an X-BOSS used in conjunction with an X-ray telescope
can lead to tremendous improvements in angular resolution.  
The trend of increasing the effective area of future X-ray telescopes at the
expense of angular resolution (Table~\ref{tab:xray-telescopes}) meshes perfectly
with the benefits gained by including an X-BOSS in the mission.  Indeed, an
X-ray telescope to be used with an X-BOSS is treated as a light bucket with
all the resolving power coming from the X-BOSS occulting the source.  Thus
an X-BOSS is an excellent addition to an X-ray telescope mission,
particularly one at L2, such as Constellation X where sub-milliarcsecond
resolution can be attained for a wide range of sources.

For the Chandra X-ray telescope we found that moderate improvements in
angular resolution over an appreciable fraction of the sky can be achieved
through the use of an X-BOSS\@.  Similarly an X-BOSS employed in
conjunction with XMM would provide tremendous improvements in the angular
resolution that XMM could achieve allowing XMM to have angular resolution
comparable to Chandra. An X-BOSS launched for use with Chandra or XMM would
also provide an important test bed for the technology to be used with
future missions.

\acknowledgements
The authors would like particularly to thank 
Art Chmielewski for financial and other support,
A. Babul and M. Dragovan for many useful comments and suggestions,
N. Choudhuri for helpful suggestions on statistical tests, and Paul
Gorenstein and William Zhang for comments on a preliminary version of the 
manuscript.
This work was supported by 
a CAREER grant to GDS from the National Science Foundation,
a DOE grant to the theoretical particle and astrophysics group at CWRU, 
by a grant from NASA's Jet Propulsion Laboratory,
and by funds from CWRU.

\begin{deluxetable}{lccl}
\tablecaption{Properties of existing and planned X-ray
  telescopes.\label{tab:xray-telescopes}}

\tablehead{
  \colhead{Satellite} & \colhead{Effective} & \colhead{Angular} &
  \colhead{Orbit} \\ 
  \colhead{Name} & \colhead{Area at $1\keV$} & \colhead{Resolution} &
  \colhead{}\\ 
  \colhead{} & \colhead{(cm$^2$)} & \colhead{(arcsecond)} & \colhead {}
  }
\startdata
Chandra (AXAF)\tablenotemark{a}  & \phn\phn\phn\phm{,}700 & 0.5 & eccentric
high Earth orbit \\
XMM             & \phn\phn2,000 & 15         & eccentric high Earth orbit \\
Astro-E         & \phn\phn1,200       &    90        & low Earth orbit \\
HETE-II\tablenotemark{b} & \phn\phn\phn\phm{,}350 & 660 & low Earth orbit \\
Constellation X & \phn30,000\tablenotemark{c} &  15        & L2 halo orbit \\
Xeus -- Phase I &   \phn60,000    &      2        & low Earth orbit \\
\phantom{Xeus}~-- Phase II&   300,000   &      2       & low Earth orbit \\
\enddata
\tablenotetext{a}{Effective area is for the AXAF CCD Imaging Spectrometer
  (ACIS)\@.  Angular resolution is for the High Resolution Camera (HRC)\@.}
\tablenotetext{b}{These values are for the wide field X-ray monitor (WXM)
  instrument.  The quoted effective area is for $2\keV$ X-rays.}
\tablenotetext{c}{Total effective area for all modules.}
\tablerefs{\\
Chandra: http://asc.harvard.edu/\\
XMM:  http://xmm.vilspa.esa.es/\\
HETE-II: http://space.mit.edu/HETE/\\
Astro-E: http://heasarc.gsfc.nasa.gov/docs/astroe/overview.html\\
Constellation X: http://constellation.gsfc.nasa.gov/\\
Xeus: http://astro.estec.esa.nl/SA-general/Projects/XEUS/web/mission.html
}

\end{deluxetable}

\end{document}